\documentclass[runningheads]{llncs}
\usepackage{graphicx}
\usepackage{hyperref}

\usepackage[cmex10]{amsmath}
\usepackage[]{amssymb, amsfonts}%
\usepackage{algorithm}
\usepackage{algpseudocode}
\algrenewcommand{\algorithmicrequire}{\textbf{Input:}}
\algrenewcommand{\algorithmicensure}{\textbf{Output:}}

\usepackage[caption=false,font=footnotesize]{subfig}

\graphicspath{Figures}

\newtheorem{ex}{Example}

\usepackage{xcolor}
\usepackage{xspace}
\usepackage{marginnote}
\usepackage[normalem]{ulem}

\newcounter{mnotes}
\newcommand{\cothomas}[1]{
    \stepcounter{mnotes}
    \def\@currentlabel{\themnotes}
    \raisebox{.1ex}{\framebox{\tiny\rm\color{blue}\themnotes}}
    \marginnote{\framebox{\rm\color{blue}\themnotes}:(TG) #1}[3cm]
}
\newcommand{\cowenbin}[1]{
    \stepcounter{mnotes}
    \def\@currentlabel{\themnotes}
    \raisebox{.1ex}{\framebox{\tiny\rm\color{orange}\themnotes}}
    \marginnote{\framebox{\rm\color{orange}\themnotes}:(WZ) #1}[3cm]
}

\newcommand{\ie}{\emph{i.e.}\xspace}
\newcommand{\cf}{\emph{cf.}\xspace}
\newcommand{\eg}{\emph{e.g.}\xspace}
\newcommand{\al}{\emph{al.}\xspace}

\newcommand{\IncSeq}{{\sc IncSeq}\xspace}

\begin{document}
\title{Incremental Mining of Frequent Serial Episodes Considering Multiple Occurrences}

\titlerunning{Incremental mining of frequent serial episodes}        % if too long for running head

\author{Thomas Guyet\inst{1} \and Wenbin Zhang\inst{2} \and Albert Bifet\inst{3,4}}

\authorrunning{T. Guyet, W. Zhang and A. Bifet} % if too long for running head

%%%% list of authors for the TOC (use if author list has to be modified)
%\tocauthor{T. Guyet et al.}

%Inria, 56 rue Niels Bohr, F-69100 Villeurbanne, France\\

\institute{Inria, Lyon Center, France\\
              \email{thomas.guyet@inria.fr}
           \and
           		Carnegie Mellon University, United States\\
           		\email{wenbinzhang@cmu.edu}
           \and
              University of Waikato, New Zealand \\
            \and
            	LTCI, Telecom Paris, Institut Polytechnique de Paris, France\\
            	\email{albert.bifet@waikato.ac.nz}
}

\maketitle              % typeset the header of the contribution
\setcounter{footnote}{0}
%
% \vspace{-0.5cm}
\begin{abstract}
The need to analyze information from streams arises in a variety of applications. One of its fundamental research directions is to mine sequential patterns over data streams. Current studies mine series of items based on the presence of the pattern in transactions but pay no attention to the series of itemsets and their multiple occurrences. The pattern over a window of itemsets stream and their multiple occurrences, however, provides additional capability to recognize the essential characteristics of the patterns and the inter-relationships among them that are unidentifiable by the existing presence-based studies. In this paper, we study such a new sequential pattern mining problem and propose a corresponding sequential miner with novel strategies to prune the search space efficiently. Experiments on both real and synthetic data show the utility of our approach.  
\keywords{event sequence, serial episode, multiple occurrences}

\end{abstract}

\section{Introduction}

Online mining of frequent patterns over a sliding window is one of the most important tasks in data stream mining with broad applications. In this case, the data stream is made of items or itemsets that arrive continuously. The aim is then to obtain a set of evolving frequent patterns over a sliding window, in which the most recent frequent patterns as well as their evolution are available at any time for information extraction. This motivates work on mining frequent patterns over series of items based on their presence in the stream~\cite{calders2014mining,zihayat2017efficiently}. In this paper, to gain additional information from the stream, we take one step further to extract frequent sequential patterns over a stream of itemsets but also to consider their multiple occurrences in the stream.  

% This task has been addressed through the extraction of frequent itemsets considering the presence of the pattern~\cite{calders2014mining,zihayat2017efficiently}. In this article, we propose to extend these works by extracting frequent sequential patterns over a stream of itemsets. 

Mining frequent sequential patterns from a single long sequence $S$ is better known as serial episode mining \cite{Mannila1997}. Under this setting, the support of a pattern is the number of times it occurs in $S$. 
The way to enumerate the multiple occurrences of a pattern turns out to be important to have the antimonotonicity of the measure. Among the possible enumeration strategies \cite{Achar2010}, the \textit{minimal occurrences} is the most common \cite{Mannila1997} with the initial work discussed in~\cite{guyet2012incremental}. With this property, the classical breadth-first search (like PrefixSpan \cite{Pei2004}) or depth-first search algorithms (like GSP \cite{Srikant96}) can be adapted to efficiently extract the complete set of frequent sequential patterns occurring in a static sequence. However, applying such algorithms to maintain the recent frequent patterns over the stream would be intractable. In addition, start from scratch each time a new item arrives in the stream is needed, but the computation cost, in practice, is unaffordable.  

% It would require to run it from scratch each time a new item arrives in the stream. The computation cost, in practise, is affordable. 

%The computation cost would not be acceptable.

% In this paper, we adapt serial episode mining in the context of stream. To this end, we introduce a novel algorithm \IncSeq to efficiently extract frequent serial episodes over the stream. 
% To the best of our knowledge, this is the first algorithm for mining serial episodes incremental. Then, our objective is to propose an algorithm that is 
% The major contributions of this paper are: 

%{\IncSeq stands for ``{\sc Inc}remental {\sc Seq}uence''.}

To address the aforementioned challenges, this paper introduces {\sc Inc}remental {\sc Seq}uence (\IncSeq), a novel framework to efficiently extract frequent serial episodes over the stream of itemsets. To the best of our knowledge, this is the first work capable of mining series of itemsets incrementally without the need to start from scratch. To summarize, we present the following contributions:

\begin{itemize}
	\item The formalization of a new incremental sequential pattern mining problem, which counts the exact number of occurrences of sequential patterns. 
	\item A complete algorithm for incremental sequential pattern mining with efficient search space pruning.
	\item Extensive experiments on both real and synthetic datasets. 
\end{itemize}

\section{Basic Concepts and Problem Statement} \label{sec:Notations}

Suppose that we have a set of items denoted $\mathcal{E}$ 
and $<$ defines the total order on this set (\eg lexicographic order). 
An itemset $\beta=(b^i)_{i\in [m]}\subseteq\mathcal{E}$ is a sub-itemset of $\alpha=(a^i)_{i\in [n]}\subseteq\mathcal{E}$, denoted $\beta \sqsubseteq \alpha$, iff there exists a sequence of integers $1 \leq i_1 < i_2 < \cdots  < i_m \leq n$ such that $\forall k \in [m], b^k=a^{i_k}$.\footnote{$[n]$ denotes the set of the $n$ first integers $\{1, \dots, n\}$.} A \emph{sequence} $S$ is a finite ordered series of itemsets $S = \langle s_1, s_2, \dots, s_n \rangle$. A \emph{serial episode} (also called \emph{sequential pattern} or pattern for short) is a \emph{sequence}. The length of a sequential pattern $S$, denoted $|S|$, is the number of itemsets it contains. The total number of items in a pattern $S$ is denoted $\|S\|$.
$T=\langle t_1, t_2, \dots, t_m\rangle$ is a \emph{sub-sequence} of $S=\langle s_1, s_2, \dots, s_n\rangle$, denoted $T \preceq S$, iff there exists a sequence of integers $1 \leq i_1 < i_2 < \cdots  < i_m \leq n$ such that $t_k \sqsubseteq s_{i_k}$ for all $k \in [m]$.

The \textbf{minimal occurrences}~\cite{Mannila1997} of a sequential pattern $S=\langle s_1, \ldots, s_n\rangle$ in a sequence $W=\langle w_1, \ldots, w_m\rangle$, denoted $\mathcal{I}_W(S)$, is the list of $n$-tuple of positions (within $W$): 
{
\small
\begin{equation}\label{eq:Instances}
\begin{array}{lll}
\mathcal{I}_{W}(S)=\left\{ (i_j)_{j\in [n]} \in [m] \;|\right.
   & \forall j \in [n], s_j \sqsubseteq w_{i_j}, & \text{(a)} \\
   & \forall j \in [n-1],\; i_j < i_{j+1}, & \text{(b)} \\
   & \left(w_j\right)_{j\in\left[i_1+1, i_n\right]} \npreceq S, & \text{(c)} \\
   & \left(w_j\right)_{j\in\left[i_1, i_n-1\right]} \npreceq S  \left. \right\} & \text{(d)}
\end{array}
\end{equation}
}

In Equation~\ref{eq:Instances}, condition $(a)$ requires that any itemset of $S$ is a sub-itemset of an itemset of $W$, while condition $(b)$ specifies the order of itemsets of $W$ needs to respect. In addition, no itemset of $W$ can be a super-itemset of two distinct itemsets of $S$. This condition does not impose any time constraint between itemsets.
Conditions $(c)$ and $(d)$ specify minimal occurrences: if a minimal occurrence of $S$ has been identified in the interval $\left[i_1, i_n\right]$, there can not be any minimal occurrence of $S$ in a strict subinterval of $\left[i_1, i_n\right]$. For sake of simplification, ``occurrence'' denotes ``minimal occurrence'' in the remainder of this paper.

Then, the \emph{support} of a sequential pattern $S$ in sequence $W$, denoted $supp_W(S)$, is the cardinality of $\mathcal{I}_{W}(S)$, \ie $supp_W(S) = card\left(\mathcal{I}_{W}\left(S\right)\right)$. The support measure $supp_W(\cdotp)$ is anti-monotonic on the set of sequential patterns with associated partial order $\preceq$ \cite{Tatti2012}. Given a threshold $\sigma$, we say that a sequential pattern $S$ is \emph{frequent} in a stream window $W$ iff $supp_W(S) \geq \sigma$. 

Mining frequent sequential patterns \textbf{incrementally} is therefore to extract frequent sequential patterns in a sequence $W=\langle w_1, \ldots, w_m \rangle$ from the ones in $W'=\langle w_0, \ldots, w_{m-1}\rangle$. This recursively mining of frequent sequential patterns enables to mine a stream of itemsets, \ie to maintain the set of frequent sequential patterns in a window sliding over a stream of itemsets.

% The recursive application of mining frequent sequential patterns incrementally enables to mine a stream of itemsets, \ie to maintain the set of frequent sequential patterns in a window sliding over a stream of itemsets.

%\textcolor{red}{i could not find the formal definition of \IncSeq? What does it mean? Do we need to formally introduce it before use its abbreviation if it is a abbreviation?}
%\textcolor{blue}{I added the definition of \IncSeq in the introduction ... not really "formal" ... }
 
\section{Incremental Algorithm for Sequential Patterns}
\label{sec: seq}

Our proposed approach relies on representing the set of frequent sequential patterns (or patterns for short) in a tree structure inspired by the prefixing method of PSP \cite{Masseglia1998}. 
PSP represents a set of frequent sequential patterns as a tree with two types of edges: the edges representing sequentiality ($\mathcal{S}$) between itemsets and the edges representing the composition ($\mathcal{C}$) of itemsets. Masseglia et \al~\cite{Masseglia1998} showed that such representation is memory efficient.

Formally, a tree node $N$ is a 4-tuple $\langle \alpha, \mathcal{I}, \mathcal{S}, \mathcal{C}\rangle$ where: 
\begin{itemize}
 \item $\alpha=\left(a_i\right)_{i\in[n]}$ is a sequential pattern of size $n$,
 \item $\mathcal{I} = \mathcal{I}_{W}(\alpha)$, the list of minimal occurrences of $\alpha$ in $W$,
 \item $\mathcal{S}$ is the set of descendant nodes which represent patterns $\beta=\left(b_i\right)_{i\in[n+1]}$ of size $\|\alpha\|+1$ such that $\forall i \in [n], \; a_i=b_i$, %(\ie $b_{n+1}$ is a direct successor of $a_n$),
 \item $\mathcal{C}$ is the set of descendant nodes which represent patterns $\beta=\left(b_i\right)_{i\in[n]}$ of size $\|\alpha\|+1$ such that $\forall i \in [n-1], \; a_i=b_i$, $a_n \sqsubseteq b_n$ and $\forall j<|a_n|,\; a_n^j <b_n^{|a_n|+1}$, (\ie itemset $b_n$ extends itemset $a_n$ with the item $b_n^{|a_n|+1}$). 
\end{itemize}

A tree of frequent patterns, denoted $\mathcal{A}_{\sigma}(W)$, represents all patterns of $W$ having a support greater than $\sigma$.
The root node of a prefix tree is a node of the form $\langle \{\}, \emptyset, \mathcal{S}, \mathcal{C}\rangle$.

\begin{figure}[!htb]
 \centering
 \includegraphics[width=\textwidth]{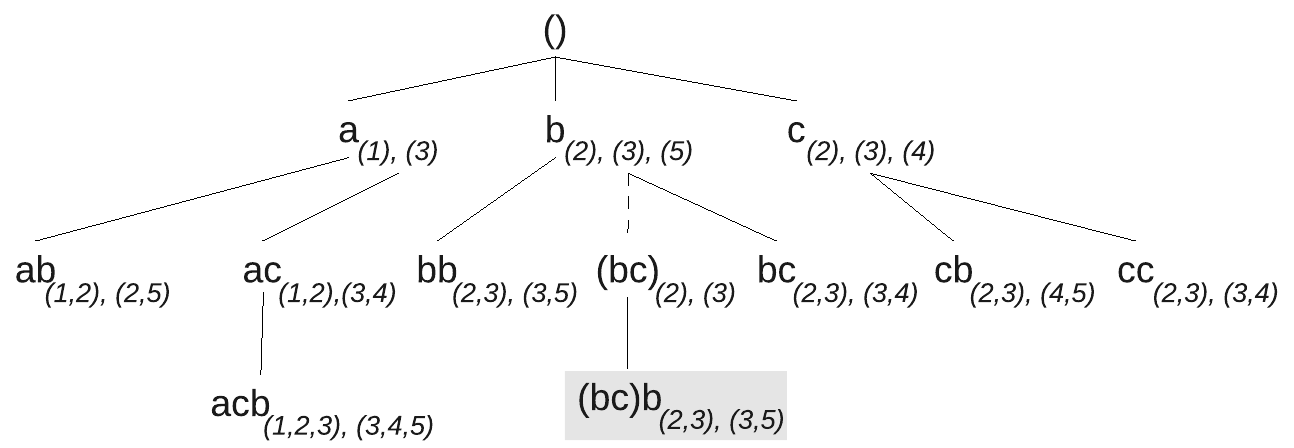}
  \caption{Example of a tree of frequent sequential patterns ($\sigma=2$) }
 \label{fig:ArbrePSP}
\end{figure}

Let $N$ be a node of $\mathcal{A}_{\sigma}(W)$. The subtree rooted at node $N$ represents the tree composed of all descendants of $N$ (including $N$). Owing to the anti-monotonicity property, we know that if a node has a support greater than or equal to $\sigma$ then all its ancestors are frequent sequential patterns in $W$. In addition, each node -- apart from the root -- has a single parent. This ensures that a recursive processing of a PSP tree is complete and non-redundant. Figure~\ref{fig:ArbrePSP} exemplifies the frequent PSP tree representation followed by its corresponding illustration. 

\begin{ex}
\label{ex:arbre}
Let $W=\langle a(bc)(abc)cb\rangle$ and $\sigma=2$. Figure \ref{fig:ArbrePSP} shows the tree $\mathcal{A}_{\sigma}(W)$. Solid lines indicate membership in the set $\mathcal{S}$ (Succession in the sequential pattern), while the dotted lines indicate membership in the set $\mathcal{C}$ (Composition with the last itemset). The node $(bc)b$, highlighted in gray, has the pattern node $(bc)$ as parent, since $ (bc) b$ is obtained by concatenating $b$ to $(bc)$. The parent node of $(bc)$ is $(b)$ and is obtained by itemset composition (dotted line). At each node of Figure \ref{fig:ArbrePSP}, the list of minimal occurrences is displayed in the index. For example, the pattern $(bc) c$ has two occurrences: $\mathcal{I}(\langle (bc)c\rangle)=\{(2,3),\;(3,5)\}$. 
\end{ex}

\subsection{Illustration of the Algorithm}

The incremental process aims at updating the tree of frequent patterns with respect to the most recent window of the stream and determining which patterns are frequent. The arrival of a new itemset in the stream triggers two steps: (1) the deletion of occurrences related to the first itemset in the window; (2) the addition of patterns and occurrences related to the new incoming itemset. The addition step incurs the majority of computational load involving three substeps: (i) merging sub-itemsets of the new itemset into the current tree, (ii) completing the lists of occurrences, and (iii) pruning nodes of non-frequent patterns. Our approach therefore performs the deletion step prior to the addition of a new itemset in order to reduce the size of the tree before the computational expensive merging and completion substeps.

\begin{figure}[t]
 \centering
 \includegraphics[width=\textwidth]{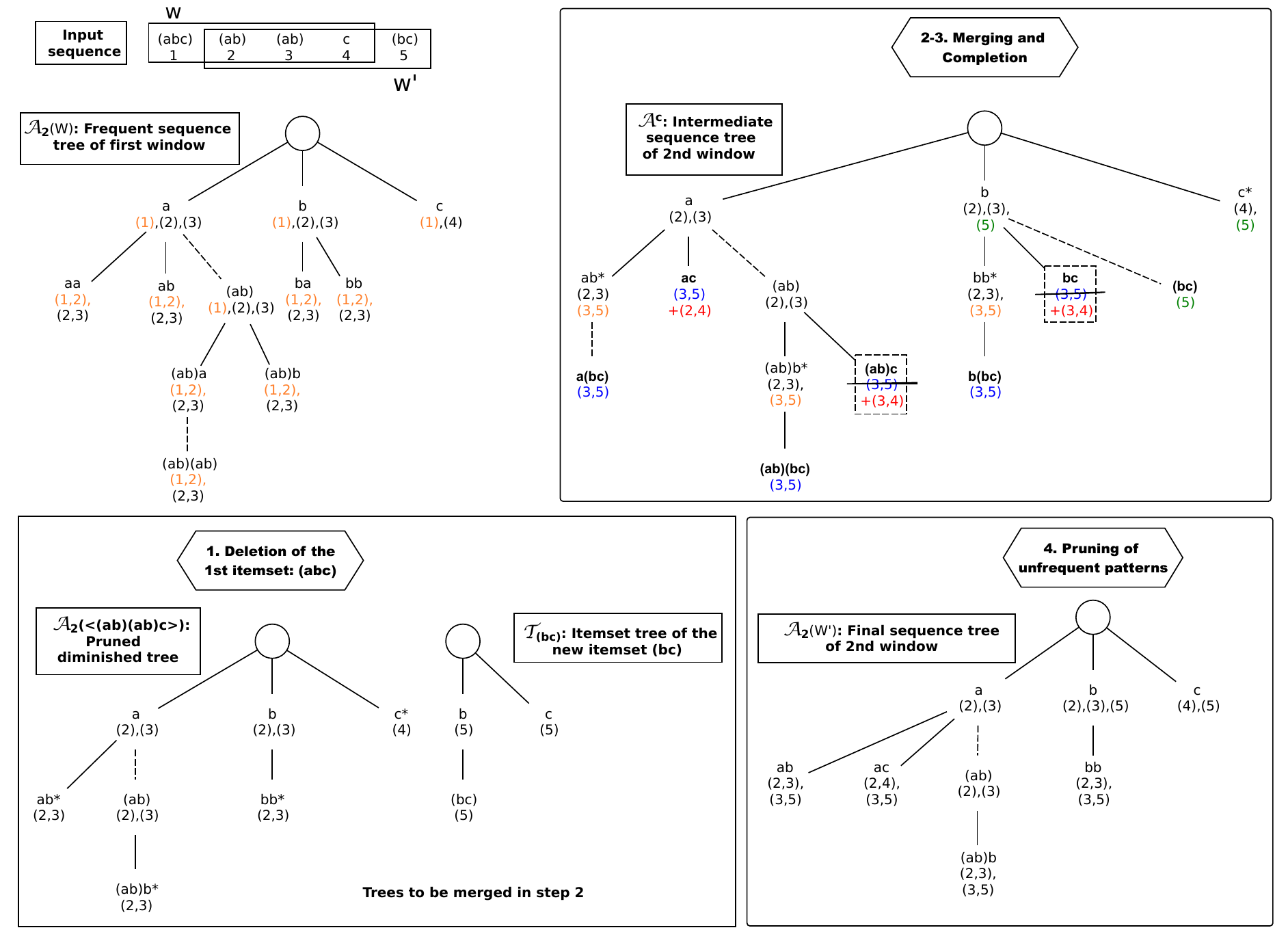}
 \caption{Successive steps for updating the tree of frequent patterns upon the arrival of itemset $(bc)$ in the window  $W=\langle (abc)(ab)(ab)c\rangle$.}
 \label{fig:IllustrationEtapes}
\end{figure}

Let us consider the window $W=\langle (abc)(ab)(ab)c\rangle$ of length 4, at position~1 of the stream. Assume that $\mathcal{A}_{2}(W)$, \ie the tree of patterns with support greater than 2, has already been built. The following steps transform the tree of frequent patterns $\mathcal{A}_{2}(W)$ into the tree $\mathcal{A}_{2}(W')$ upon the arrival of the new itemset $(bc)$. These steps are illustrated in Figure \ref{fig:IllustrationEtapes} and detailed in the following.

\textbf{1. Deletion of the first itemset}: all occurrences starting at the first (oldest) position of the window (orange occurrences at position 1 in the example) are deleted. Then, patterns having a number of occurrences lower than $\sigma = 2$ are deleted from the tree. The result is the tree $\mathcal{A}_{2}(\langle (ab)(ab)c \rangle)$ where $a$, $(ab)$, $b$ are frequent. Quasi-frequent patterns (marked with asterisk in the example) are not frequent but may become frequent as they have a support equals to $\sigma-1$ and they are ended by an item present in the new itemset, \ie $(bc)$. Such nodes are kept in the frequent tree with their occurrences as the following completion step (see below) is not necessary for them.

\textbf{2. Merging the new current itemset $(bc)$ with every node of the tree of patterns}: this step generates all the new candidate patterns of the new window. Intuitively, a pattern is a new candidate (\ie potentially frequent) only if it is the concatenation of a sub-itemset of $(bc)$ to a frequent pattern of $\langle (ab)(ab)c\rangle$.
In the tree representation of frequent patterns, this concatenation can be seen as extending each node of $\mathcal{A}_{2}(\langle (ab)(ab)c\rangle)$ with the itemset tree $\mathcal{T}_{(bc)}$ representing all sub-itemsets of $(bc)$.

In Figure \ref{fig:IllustrationEtapes}, the tree $\mathcal{T}_{(bc)}$ is merged with the four non-quasi-frequent nodes of $\mathcal{A}_{2}(\langle (ab)(ab)c\rangle)$:
\begin{itemize}
  \item with the root node (green occurrences): all subsequences of $(bc)$ become potentially frequent.
  \item with the nodes $a$, $(ab)$, $b$ (blue occurrences): all patterns starting with one of these three patterns (frequent in $\langle (ab)(ab)c\rangle$) and followed by a sub-itemset of $(bc)$ become potentially frequent.
\end{itemize}

We call this procedure ``tree merging'' because if a node already exists in the tree (\eg node $(b)$), the occurrences related to the new itemset are added to the list of existing occurrences. The list of occurrences of $(b)$ becomes $\{(2), (3), (5)\}$. We know that each of these nodes holds all the occurrences of the associated pattern in $W'$. New nodes are noted in bold face in the frequent tree after the merging step in Figure \ref{fig:IllustrationEtapes}. Each of these new nodes of $\mathcal{A}^{f}$, \eg the node $(bc)$, has an occurrence list consisting of only one occurrence of a sub-itemset of $(bc)$. Quasi-frequent nodes (nodes marked with the asterisk) are not merged with the itemset tree $\mathcal{T}_{(bc)}$. Their occurrence lists are simply updated when needed.

\textbf {3. Completion of occurrences' lists}: Exclusively for new candidate nodes, it is necessary to scan the window $W'$ once again to build the complete list of occurrences of a pattern. For example, the node $ab$ is associated with the list $\{(3,5)\}$. This list must be completed with the occurrences of $ab$ in the previous window ($\{(2,3)\}$). As $\langle ab\rangle$ was unfrequent in $W$, we must retrieve their occurrences. 
Red occurrences of the tree $\mathcal{A}^{c}$ in Figure \ref{fig:IllustrationEtapes} show the occurrences added by completion.

\textbf {4. Pruning non-frequent patterns}:  $\mathcal{A}^{c}$, the tree obtained after completion, contains new candidate patterns with complete lists of occurrences. The last step removes patterns with an occurrences' list of size strictly lower than $\sigma=2$ yielding the tree $\mathcal{A}_{2}(W')$.

\begin{algorithm}[!htbp]
	\caption{{\sc Merging}: merging the itemset tree $\mathcal{T}$ with every node of the tree of patterns $\mathcal{A}$.}
	\label{algo:Fusion}
	\begin{algorithmic}[1]
		%\scriptsize
		\Function{Merging}{$\mathcal{A}$, $\mathcal{T}$}
		\State $\mathcal{T}' \gets \mathcal{T}$
		\For{ $N \in \mathcal{A}$}
		\For{ $n \in \mathcal{T}'$} \Comment{Prefixing $\mathcal{T}'$}
		\State $n.\alpha = N.\alpha \oplus n.\alpha$ \Comment{Prefixing the pattern with $N.\alpha$}
		\ForAll{$I\in n.\mathcal{I}$}
		\Comment{Prefixing all occurrences}
		\State $I=d \cup I$\Comment{$d$ is the last element of $N.\mathcal{I}$}
		\EndFor
		\EndFor
		\State \Call{RecMerge}{$\mathcal{T}'$, $N$} \Comment{Recursive merging of $\mathcal{T}'$ with nodes $N$ of $\mathcal{A}$}
		\EndFor
		\State \Return{$\mathcal{A}$}
		\EndFunction
	\end{algorithmic}
\end{algorithm}

\subsection{Merging a Tree of an Itemset into a Tree of Frequent Patterns} 
Now, we detail the merging step which integrates 
the itemset tree $\mathcal{T}$ into the pattern tree $\mathcal{A}$. Then, we explain the completion of occurrences.

Algorithm \ref{algo:Fusion} describes how the itemset tree $\mathcal{T}$ is merged with every node of the frequent patterns tree  $\mathcal{A}$. It consists of two main steps:
\begin{itemize}
  \item prefixing the itemset tree $\mathcal{T}$ with the pattern of node $N$,
  \item recursively merging the prefixed $\mathcal{T}$ with descendants of node $N$ (\cf Algorithm \ref{algo:RecMerge}).
\end{itemize}

Let $N.\alpha$ be the pattern associated with a node $N$ from the tree of patterns $\mathcal{A}$ and $N.\mathcal{I}$ be the list of occurrences associated with $N$. For each node $N$ of $\mathcal{A}$, the itemset tree $\mathcal{T}$ is first prefixed by $N$: on the one hand, the patterns of each node of $\mathcal{T}$ are prefixed by $N.\alpha$; on the other hand, all occurrences of $\mathcal{T}$ are prefixed by the last occurrence of $N.\mathcal{I}$.
Using the last occurrence in $N.\mathcal{I}$ enforces the third property (see eq. \ref{eq:Instances}). 

\begin{algorithm}[!htbp]
\caption{{\sc RecMerge}: recursively merging the prefixed itemset tree $\mathcal{T}$ with a node of $\mathcal{A}$}
\label{algo:RecMerge}
\begin{algorithmic}[1]
%\scriptsize
\Require{$n$: itemset node tree, $N$: node of the tree of patterns to be merged with $n$ and such that $n.\alpha=N.\alpha$}
\Function{RecMerge}{$n$, $N$}

\State $N.\mathcal{I} \gets N.\mathcal{I}\cup n.\mathcal{I}$ \Comment{Merging lists of occurrences} \label{algo:RecMerge:fusioninstances} 
\For{ $s_N \in N.\mathcal{S} \cup N.\mathcal{C}$} \Comment{Recursion}\label{algo:RecMerge:S_debut}
  \For{ $s_n \in n.\mathcal{S} \cup n.\mathcal{C}$}
    \If{ $s_N.\alpha = s_n.\alpha$ }
      \State $found \gets \texttt{True} $ 
      \State \Call{RecMerge}{$s_n$, $s_N$}
    \EndIf
  \EndFor
  \If{ not $found$ }
    \If{ $s_n \in n.\mathcal{S}$ }
      \State $N.\mathcal{S} \gets N.\mathcal{S}\cup\{$\Call{Copy}{$s_n$}$\}$
    \Else
      \State $N.\mathcal{C} \gets N.\mathcal{C}\cup\{$\Call{Copy}{$s_n$}$\}$
    \EndIf
  \EndIf\label{algo:RecMerge:S_fin}
\EndFor
\EndFunction
\end{algorithmic}
\end{algorithm}

In a second step, the algorithm recursively merges the root of the itemset tree $\mathcal{T}$ prefixed by $N$. 
Algorithm \ref{algo:RecMerge} details this merging operation. We first need to make sure that $n.\alpha = N.\alpha$ to verify that the two nodes represent the same pattern.
At line \ref{algo:RecMerge:fusioninstances}, occurrences of nodes $n$ and $N$ are merged. By construction of the new occurrence, the conditions of Eq. \ref{eq:Instances} are satisfied.
Then, the descendants of $n$ are processed recursively. For each node of $n.\mathcal{S}$ (resp. $n.\mathcal{C}$), we search a node $s_n$ in $N.\mathcal{S}$ (resp. $N.\mathcal{C}$) such that these nodes represent the same pattern. If such a node is found, then the function {\sc RecMerge} is recursively applied. Otherwise, a copy of the entire subtree of $s_n$ is added to $n.\mathcal{S}$ (resp. $n.\mathcal{C}$).

\subsection{Completion of a List of Occurrences}
When a new pattern is introduced in the tree, it means that it was unfrequent in the previous window, but there might exist occurrences of this pattern. They were simply not stored in the tree (except quasi-frequent patterns). For example, in Figure \ref{fig:IllustrationEtapes}, the pattern $\langle bc \rangle$ (node surrounded by a dotted line square) is not frequent in $W$ and is not present in the frequent patterns tree $\mathcal{A}_{2}(W)$. 
However, after the arrival of itemset $(bc)$ the pattern $\langle bc \rangle$ may become frequent in $W$. Thus, it is necessary to scan $W'$ to retrieve all occurrences of $\langle bc \rangle$ to compute its frequency.

The completion algorithm is applied exclusively to the nodes newly introduced in the tree. While ensuring the completeness, this method reduces the number of completions. In addition, to make the completion efficient, the occurrences of a pattern $\beta$ is recursively constructed from the occurrences of its direct parent along the following principles:
\begin {itemize}
  \item each occurrence $I=(i_1, \dots,i_{|\delta|})$ of a pattern $\delta$ obtained by adding an item $e$ to the last itemset of $\beta$ (composition) are necessarily occurrences of $\beta$, thus the algorithm tests whether $e$ is included in the itemset $w_{i_{|\delta|}}$.
  \item each occurrence $I=(i_1, \dots,i_{|\epsilon|})$ of a pattern $\epsilon$, obtained by adding an itemset $e$ to $\beta$ (succession), are necessarily constructed by adding the element $i_{|\epsilon|}$ to an occurrence of $\beta$, thus the algorithm browses a sub-sequence of $W'$ to test the presence of $e$.
\end {itemize}

For succession nodes, the completion scans only the sub-sequence of $W'$ composed of the itemsets between $i_{|\beta|}+1$ and $j_{|\beta|-1}$, where $J=(j_1,\dots,j_{|\beta|})$ is the occurrence after $I$ in the list of occurrences of $\beta$.

As an example, on the tree $\mathcal{A}^{c}$ in Figure \ref{fig:IllustrationEtapes}, the occurrences of $\langle bc \rangle$ is $\mathcal{I}(\langle bc \rangle)=\{(3,5)\}$. This occurrence has been obtained during the merging step by adding the element $5$ to the occurrence $(3)$ of pattern $\langle b \rangle$.
An occurrence of $\mathcal{I}(\langle bc \rangle)$ is the successor of one of the occurrences of $\langle b \rangle$: $\mathcal{I}(\langle b \rangle)=\{(2), (3), (5)\}$.
To complete occurrence $(3)$ from $\mathcal{I}(\langle b \rangle)$, the algorithm looks for one $c$ in $W'$ at a position between $3$ ($=2+1$) and the beginning of the third occurrence of $\mathcal{I}(\langle b \rangle)$, \ie $5$. Here, occurrence $(3,4)$ is found. But it is a sub-sequence of an existing occurrence $(3,5)$. Due to the definition of \textit{minimal} occurrences (eq. \ref{eq:Instances}), $(3,5)$ is deleted. The same for pattern $(ab)c$ (the other node surrounded by a dotted line square). It is not possible to complete occurrence $(2)$ of $\mathcal{I}(\langle b \rangle)$ because there is no $c$ in the itemset at position $3$ (the only possible itemset between the occurrence of $\langle b \rangle$ at position ($2$) and the next occurrence in $\mathcal{I}(\langle bc \rangle)$).

It is worth mentioning that the proposed algorithm is complete. Specifically, in a streaming context which applies recursively the incremental mining process, it extracts all the frequent sequential patterns for each sliding window of the stream.

\section{Experiments and Results} \label{sec:Results}

The objective of our experiments is to show that the proposed algorithm is an efficient strategy for mining sequential patterns incrementally. 
More specifically, we would like to assess the space and time efficiency of the proposal compared to a $Batch$ approach, \ie a strategy that does not exploit the incremental changes of the window. The $Batch$ algorithm is based on PrefixSpan and uses the PSP tree structure. It rebuilds the entire tree $\mathcal{A}_{\sigma}(W)$ for each consecutive window of size $ws$ on the data stream. 
To the best of our knowledge, there is no state-of-the-art competitor for this task.

It is worth noticing that the two approaches are complete and thus extract the exact same sets of patterns. For this reason, we do not discuss the algorithm outputs but only their efficiency. 

In a first experiment, we present the result on synthetic data which have been widely used to evaluate the efficiency of sequential pattern mining algorithms. 
As the purely random nature of this data does not mimic the characteristics of true datasets (with less balanced itemset occurrences or with the presence of significant patterns), we supplement this experiment with an experiment on a real dataset. This dataset also illustrates the practical value and additional information gained by addressing the newly formulated sequential pattern mining problem of this work. 

% This dataset also illustrate a practical use case where sequential pattern over a data stream of itemsets considering multiple occurrence.

The algorithms were implemented in C++ and ran on a single core. The source code, synthetic datasets and benchmarks scripts are available online\footnote{\url{https://gitlab.inria.fr/tguyet/seqstreamminer}}.

\subsection{Experiments on Synthetic Data}
In this section, we evaluate \IncSeq against $Batch$ on synthetic datasets generated in the same way as the IBM quest data generator. Specifically, at each sequence position, an item is present with a probability of 3\%, thus yielding a random sequence of itemsets. The length of the sequence simulating the stream is 1000 times of the windows size, which requires the incremental algorithm to be recursively called 1000 times in a run. The item vocabulary size, $card(\mathcal{E})$, is set to 40. Then, the average number of items per itemset is $1.2$. The experiments were conducted by varying the parameters $ws$ (window size from 80 to 300) and $\sigma$ (minimal support from 3 to 10 occurrences) on 5 different datasets per configuration. The results reported are the average results of all the experiments.

\begin{figure}[!htbp]
\centering
\subfloat[]{
	 \includegraphics[width=.5\textwidth, height=0.3\textheight]{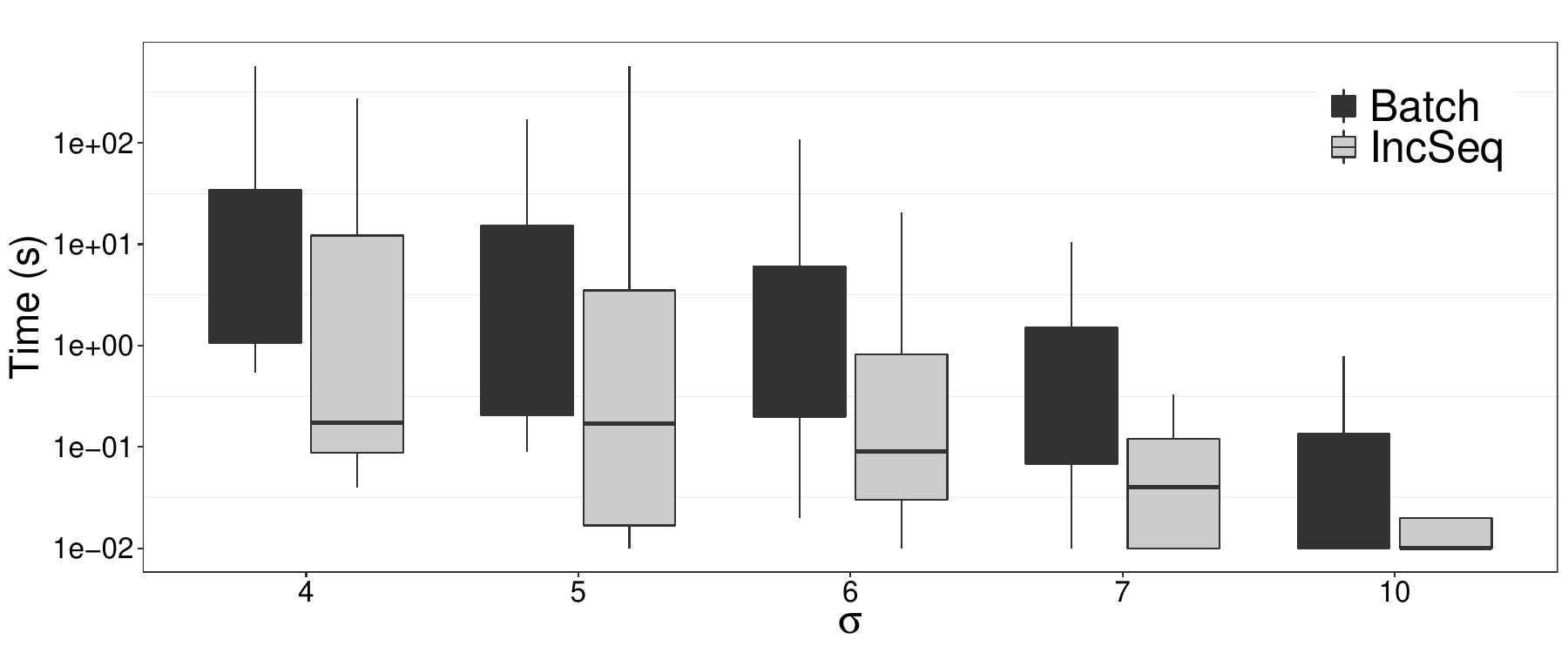}
	}%\\  \vspace{-10pt}
\subfloat[]{
	 \includegraphics[width=.5\textwidth, height=0.3\textheight]{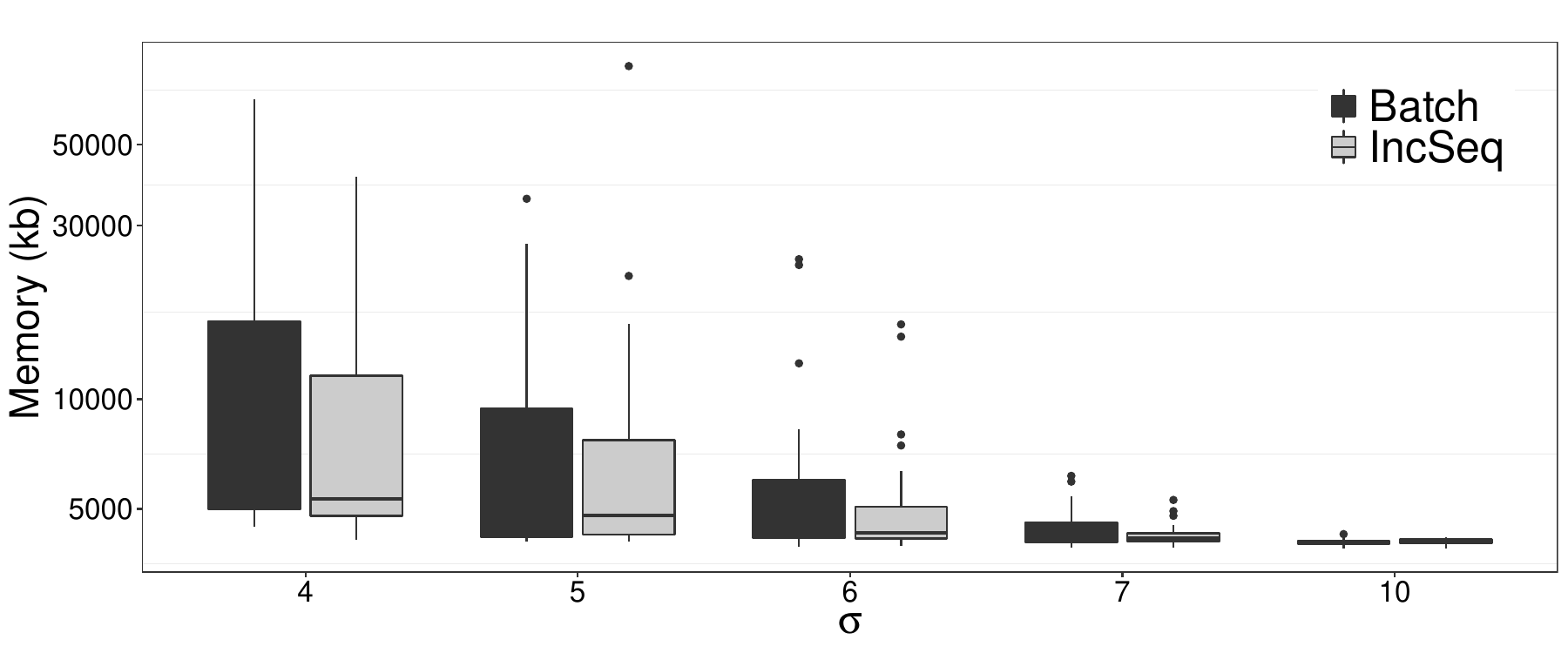}
	}\\  \vspace{-10pt}
\subfloat[]{
	 \includegraphics[width=.5\textwidth, height=0.3\textheight]{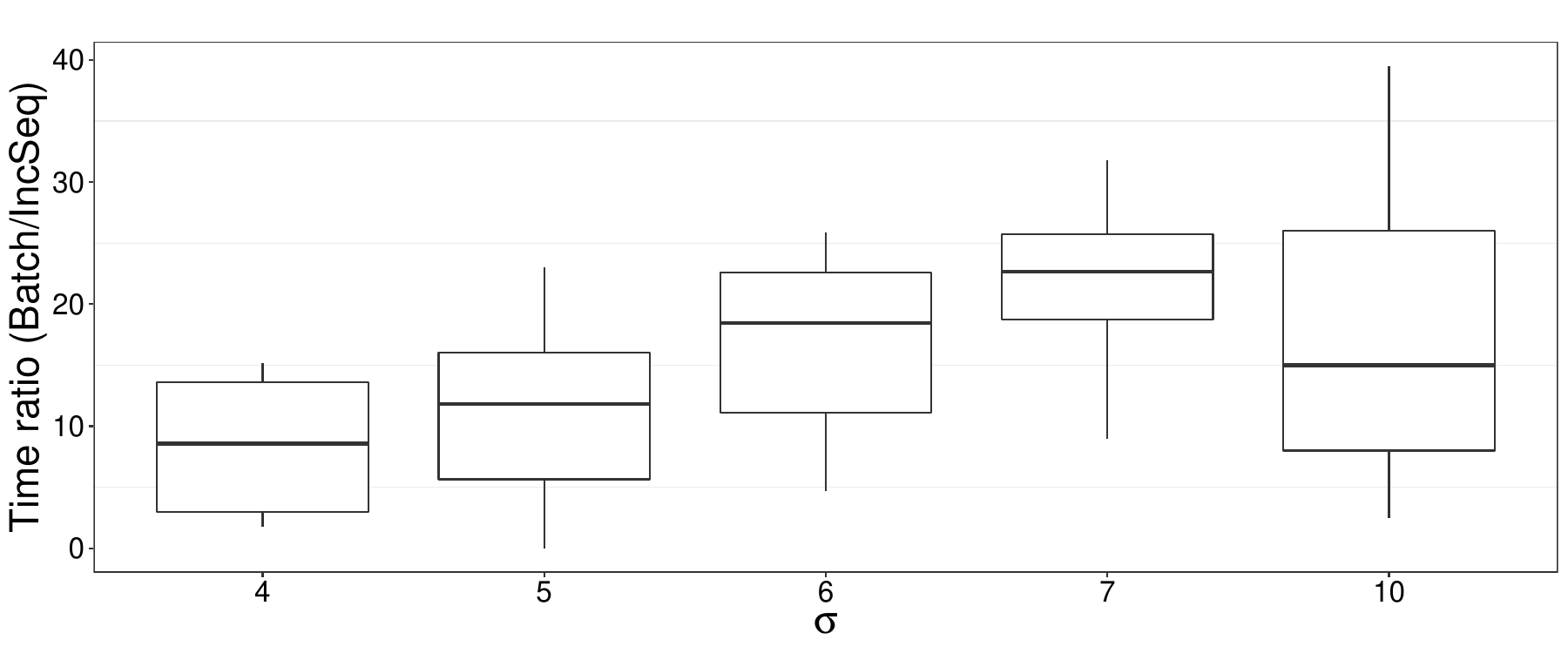}
	}%\\  \vspace{-10pt}
\subfloat[]{
	 \includegraphics[width=.5\textwidth, height=0.3\textheight]{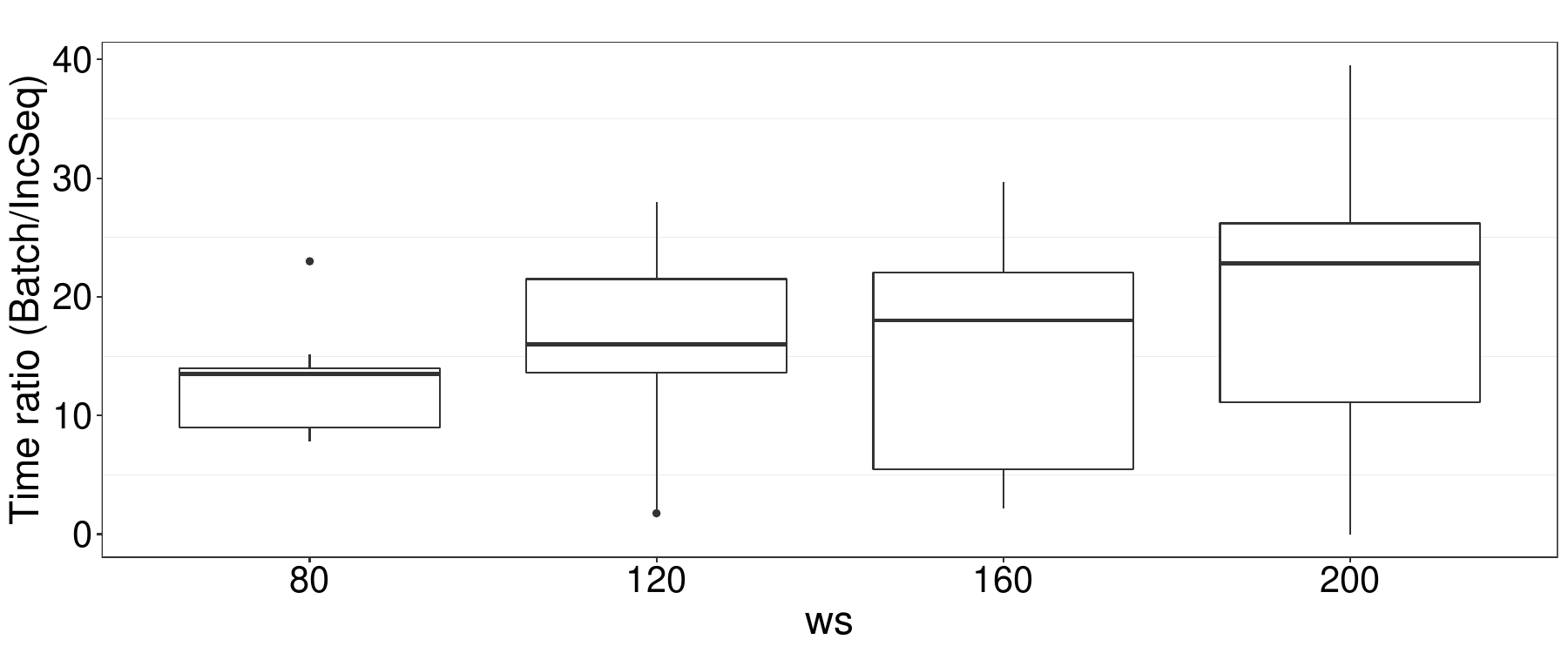}
	}
\caption{Comparison of processing time (logarithmic scale) and memory usage with respect to the support threshold $\sigma$ (with $ws<25$) and the size of the sliding window $ws$. (c) and (d) represent the respective computing time ratio of Batch to \IncSeq on the same dataset.}
\label{fig:results}
\end{figure}

%\begin{figure}[t]
%\centering
%\subfloat[]{
% \includegraphics[width=.46\textwidth, height=3.1cm]{./Results/timeVSsigma}
%}%\\  \vspace{-10pt}
%\subfloat[]{
% \includegraphics[width=.46\textwidth, height=3.1cm]{./Results/memVSsigma}
%}\\  \vspace{-10pt}
%\subfloat[]{
% \includegraphics[width=.46\textwidth, height=3.1cm]{./Results/timeratioVSsigma}
%}%\\  \vspace{-10pt}
%\subfloat[]{
% \includegraphics[width=.46\textwidth, height=3.1cm]{./Results/timeratioVSws}
%}
%\caption{Comparison of processing time (logarithmic scale) and memory usage with respect to the support threshold $\sigma$ (with $ws<25$) and the size of the sliding window $ws$. (c) and (d) represent the respective computing time ratio of Batch to \IncSeq on the same dataset.}
%\label{fig:results}
%\end{figure}

Figure \ref{fig:results}-(a) illustrates the execution time with respect to $\sigma$. As one can see, the execution time grows exponentially when $\sigma$ decreases. Note that a timeout is set as 10 minutes. For more time-consuming mining tasks (with low $\sigma$), Batch failed 17 times before a successful completion of the mining process, while \IncSeq failed 16 times.  
It is also clear that \IncSeq, on average, is an order of magnitude faster than $Batch$. To further assess the superior efficiency of \IncSeq on mining various sizes of window, Figure \ref{fig:results}-(c) and (d) provide the execution time ratio between \IncSeq and $Batch$ with respect to $\sigma$ and $ws$, respectively. As one can see, \IncSeq dominates $Batch$ by 10 to 20 times faster in processing time when $\sigma$ and $ws$ increase. 
The different drop for $\sigma=10$ because the number of frequent patterns is closed to zero. Thus, the computing times are very low for the two approaches. % The tree is very small Then mining task turns out to be trivial for both approaches.

%\textcolor{red}{what does ``trivial'' mean here? also what does "The different drop" mean? difference dimishes?}
%\textcolor{blue}{I changed the sentence. There is no pattern to extract ... the algorithms stop very early.}

Figure \ref{fig:results}-(b) additionally shows the memory usage of the two approaches. As expected, the two approaches are comparable in terms of the memory usage as the required memory is mainly to store the frequent sequential patterns and the two approaches induce identical trees. We also observe that the memory requirement depends upon $\sigma$ as the lower $\sigma$ the more frequent patterns. Ensuring memory efficiency is also an essential prerequisite for sequential pattern mining, our proposed method therefore enjoys the advantage of mining sequential patterns with reduced time at no extra memory cost. 

\subsection{Experiments on Smart Electrical Meter Data}
\label{sec:result_edf}

\begin{figure}[!htbp]
\centering
 \includegraphics[width=.49\textwidth, height=0.25\textheight]{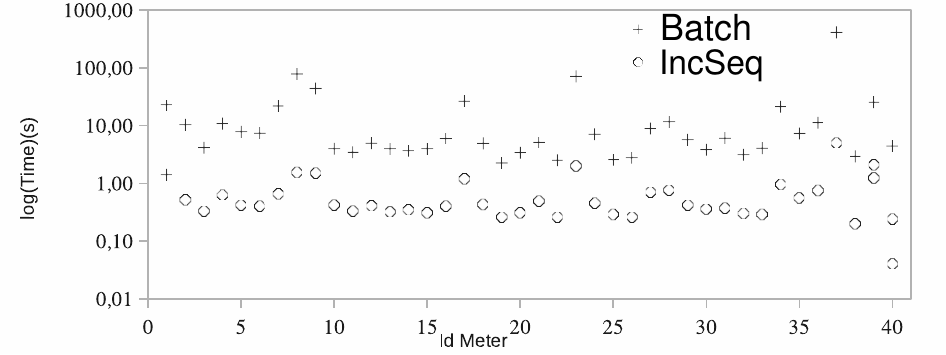}
 \includegraphics[width=.49\textwidth, height=0.25\textheight]{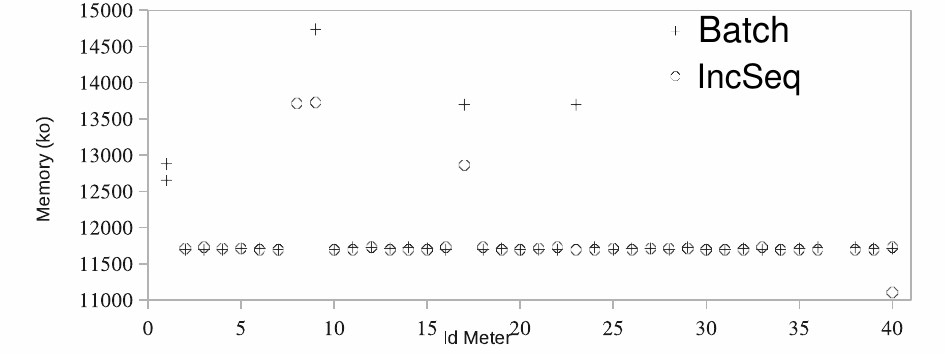}
\caption {Comparison of computation time (left) and memory usage (right) when mining the power consumption streams.}
\label {fig:results_edf}
\end {figure}

We also conducted experiments on real smart electrical meter data. Smart electrical meters record the power consumption of an individual or company in intervals of 30 min and communicate that ``instant'' information to the electric provider for monitoring and billing purposes.
The aim of smart meters is to better anticipate the high consumption of a distribution sector by awarding a consumption profile to each meter, \ie a dynamic model of changes in consumption. As consumption profiles depend on the period of year (seasons, holidays), week (weekdays, weekends) or day and are unpredictable for medium to long-term consumption, we employ \IncSeq and $Batch$ to extract the dynamic online profiles of short-term consumption of the meters.

The annual series of instantaneous consumption is a flow of about 18,000 values. We use the SAX algorithm \cite{Lin03} for discretizing the consumption values. A vocabulary size of $|\mathcal{E}|=14$ and a window aggregation of $PAA=24$ have been chosen.
The consumption profile of a smart meter at time $ t $ is the set of frequent consumption patterns during the period $[t-w,t]$ (sliding window of predefined size $ w = 28$ itemsets, \ie 2 weeks).

Figure \ref{fig:results_edf} shows the results for 40 meters. It is clear that the results obtained on the real data are consistent with those obtained on synthetic data. Specific to the real data, while most of the meters can be processed within seconds, the processing time of some meters are significantly longer (about few minutes). This disparity is attributed to the observed consumption variability. Specifically, the patterns that are more time-consuming to process are relatively constant (\eg industrial consumption) consisting of many repetitions of symbols, thus lead to a large tree depth. It is however clear that the results of real and synthetic datasets conclusively match, which suggests that our proposed method is an efficient sequential pattern miner with manageable memory cost.

\section{Related Work}

In the field of stream mining, several approaches extended frequent pattern mining in a setting similar to ours. For example, Chang et al. \cite{chang2004sliding} proposed to extract recent frequent patterns in a sliding window over a data stream, while Calders et al. \cite{calders2014mining} improved such approaches with the adaptive window size. More recently, Giacometti and Soulet \cite{giacometti2021reservoir} proposed a sampling of the pattern to improve the efficiency. Our approach focuses on more complex patterns, i.e., sequential patterns, to extract additional information with a similar streaming setup.

For sequential patterns, less efforts have been made in streaming settings~\cite{fournier2017survey}. The incremental or online sequential pattern mining algorithms in the literature address simplified problems of ours: mining frequent sequential patterns in a stream of transactions that are sequences, such as IncSPAM \cite{ho2006incremental}, or mining frequent sequential patterns in a collection of itemsets streams, such as PSP-AMS \cite{PSPAMS2018}; in both cases, the counting of sequential patterns is based on the number of transactions (resp. number of streams) in which a pattern occurs. 
%In the initial work of sequential pattern mining in streams, Chang et al. \cite{chang2005efficient} proposed eISeq in which the basic type of a sequence database is addressed. 
%In~\cite{ho2006incremental}, IncSPAM is designed to address the general sequence components considering both sequences of items and itemsets as an extension of the SPAM algorithm introduced therein. %\textcolor{red}{why do we discuss reference [6] again but not [14]? Did you try to discuss one other than [6] and [14]?}
%More recently, Koh et \al~\cite{koh2018volatility} further considers virtual concept drift, which refers to the frequency of sequence still maintains but has a decrease in its support, as apposed to real concept drift to better consider the recency constraint when discovering sequential patterns in transactional data streams. 
However, all these algorithms examine the presence of a pattern in each transaction as the pattern counting method and ignore the multiple occurrences of the pattern in a transaction. Tseng et \al \cite{Episode2016} share a similar objective, but their mining algorithms are not incremental. Their framework combines the results of episode mining by batches in a map-reduce architecture without the formal properties of \IncSeq.

Finally, our approach is also different from single-pass serial episodes mining algorithms \cite{LI2019422} whose objective is not to maintain the set of frequent serial-episodes, but is to evaluate the support of serial episodes online.

\section{Conclusion and Future Works}
\label{sec:conclusion}
Although a number of studies have developed approaches to mine sequential patterns over data streams, all of these techniques focus on a stream of items and the number of transactions that contain patterns without considering their multiple occurrences. In this work, we present our incremental algorithm based on counting the minimal occurrences of the sequential patterns over the course of itemsets stream. Experimental studies indicate the superior computational efficiency of our approach compared to the non-incremental method. In the future, we plan to further extend it by considering the condensed representation such as maximum patterns and closed patterns in the context of incremental mining. One immediate future work is to extend these results in conjunction with our previous works~\cite{zhang2022longitudinal,zhang2022kis} for fair pattern mining. A relevant avenue is to investigate the ubiquitous graph data representation~\cite{li2021time,zhang2022fairness} with unique challenges for example the independent and identically distributed (IID) data distribution.

% \vspace{-0.2cm}
% In this paper we propose a novel co-attention network for visual commonsense reasoning. We introduce an image-text fusion module to fuse information between images and text collectively, and design a novel inference module to encode commonsense among image, query and response. Extensive experiments on VCR benchmark dataset demonstrate our method significantly improves the predictions by 14.9\% in Q2A, 14.4\% in QA2R, and 32.2\% in Q2AR compared to the state of the art methods. 

%\nocite{*}
\bibliographystyle{amsplain}
\bibliography{ref}

\providecommand{\bysame}{\leavevmode\hbox to3em{\hrulefill}\thinspace}
\providecommand{\MR}{\relax\ifhmode\unskip\space\fi MR }
% \MRhref is called by the amsart/book/proc definition of \MR.
\providecommand{\MRhref}[2]{%
  \href{http://www.ams.org/mathscinet-getitem?mr=#1}{#2}
}
\providecommand{\href}[2]{#2}
\begin{thebibliography}{10}

\bibitem{Achar2010}
A.~Achar, S.~Laxman, and P.~S. Sastry, \emph{A unified view of automata-based
  algorithms for frequent episode discovery}, CoRR \textbf{abs/1007.0690}
  (2010).

\bibitem{calders2014mining}
Toon Calders, Nele Dexters, Joris~JM Gillis, and Bart Goethals, \emph{Mining
  frequent itemsets in a stream}, Information Systems \textbf{39} (2014),
  233--255.

\bibitem{chang2004sliding}
Joong~Hyuk Chang and Won~Suk Lee, \emph{A sliding window method for finding
  recently frequent itemsets over online data streams}, Journal of Information
  science and Engineering \textbf{20} (2004), no.~4, 753--762.

\bibitem{fournier2017survey}
Philippe Fournier-Viger, Jerry Chun-Wei Lin, Rage~Uday Kiran, Yun~Sing Koh, and
  Rincy Thomas, \emph{A survey of sequential pattern mining}, Data Sc. Pat.
  Reco. \textbf{1} (2017), no.~1, 54--77.

\bibitem{giacometti2021reservoir}
Arnaud Giacometti and Arnaud Soulet, \emph{Reservoir pattern sampling in data
  streams}, Proc. ECML-PKDD, 2021, pp.~337--352.

\bibitem{guyet2012incremental}
Thomas Guyet and Ren{\'e} Quiniou, \emph{Incremental mining of frequent
  sequences from a window sliding over a stream of itemsets}, Actes IAF (2012).

\bibitem{ho2006incremental}
Chin-Chuan Ho, Hua-Fu Li, Fang-Fei Kuo, and Suh-Yin Lee, \emph{Incremental
  mining of sequential patterns over a stream sliding window}, International
  Conference on Data Mining-Workshops (ICDMW), 2006, pp.~677--681.

\bibitem{PSPAMS2018}
Bijay~Prasad Jaysawal and Jen-Wei Huang, \emph{Psp-ams: Progressive mining of
  sequential patterns across multiple streams}, ACM Trans. Knowl. Discov. Data
  \textbf{13} (2018), no.~1, 1--23.

\bibitem{LI2019422}
Hui Li, Sizhe Peng, Jian Li, Jingjing Li, Jiangtao Cui, and Jianfeng Ma,
  \emph{Counting the frequency of time-constrained serial episodes in a
  streaming sequence}, Information Sciences \textbf{505} (2019), 422--439.

\bibitem{Lin03}
J.~Lin, E.~Keogh, S.~Lonardi, and B.~Chiu, \emph{A symbolic representation of
  time series, with implications for streaming algorithms}, Proceedings of the
  Workshop on Research Issues in Data Mining and Knowledge Discovery, 2003.

\bibitem{Mannila1997}
H.~Mannila, H.~Toivonen, and A.~I. Verkamo, \emph{Discovering frequent episodes
  in event sequences}, Journal of Data Mining and Knowledge Discovery
  \textbf{1} (1997), no.~3, 210--215.

\bibitem{Masseglia1998}
F.~Masseglia, F.~Cathala, and P.~Poncelet, \emph{The {PSP} approach for mining
  sequential patterns}, Proceedings of the European Symposium on Principles of
  Data Mining and Knowledge Discovery, 1998, pp.~176--184.

\bibitem{Pei2004}
J.~Pei, J.~Han, B.~Mortazavi-Asl, J.~Wang, H.~Pinto, Q.~Chen, U.~Dayal, and
  M.-C. Hsu, \emph{Mining sequential patterns by pattern-growth: the
  {PrefixSpan} approach}, Transactions on Knowledge and Data Engineering
  \textbf{16} (2004), no.~11, 1424--1440.

\bibitem{li2021time}
Tai~Le Quy, Arjun Roy, Vasileios Iosifidis, Wenbin Zhang, and Eirini Ntoutsi,
  \emph{A survey on datasets for fairness-aware machine learning}, Data Mining
  and Knowledge Discovery (2022).

\bibitem{Srikant96}
R.~Srikant and R.~Agrawal, \emph{Mining sequential patterns: Generalizations
  and performance improvements}, Proceedings of the International Conference on
  Extending Database Technology, 1996, pp.~3--17.

\bibitem{Tatti2012}
Nikolaj Tatti and Boris Cule, \emph{Mining closed strict episodes}, Data Mining
  and Knowledge Discovery \textbf{25} (2012), no.~1, 34--66.

\bibitem{Episode2016}
Jerry C.~C. Tseng, Jia-Yuan Gu, P.~F. Wang, Ching-Yu Chen, Chu-Feng Li, and
  Vincent~S. Tseng, \emph{A scalable complex event analytical system with
  incremental episode mining over data streams}, Proc. of Congress on
  Evolutionary Computation, 2016, pp.~648--655.

\bibitem{zhang2022longitudinal}
Wenbin Zhang and Jeremy Weiss, \emph{Longitudinal fairness with censorship},
  Proceedings of the AAAI Conference on Artificial Intelligence, 2022.

\bibitem{zhang2022kis}
Wenbin {Zhang} and Jeremy Weiss, \emph{Rethinking fairness: New definitions and
  algorithm for fair machine learning under uncertainty}, Knowledge and
  Information Systems (2022).

\bibitem{zhang2022fairness}
Wenbin Zhang, Jeremy~C Weiss, Shuigeng Zhou, and Toby Walsh, \emph{Fairness
  amidst non-iid graph data: A literature review}, arXiv preprint
  arXiv:2202.07170 (2022).

\bibitem{zihayat2017efficiently}
Morteza Zihayat, Cheng-Wei Wu, Aijun An, Vincent~S Tseng, and Chien Lin,
  \emph{Efficiently mining high utility sequential patterns in static and
  streaming data}, Proc. of Intelligent Data Analysis, vol.~21, 2017,
  pp.~103--135.

\end{thebibliography}
\end{document}